\pdfoutput=1

\documentclass[11pt]{article}

\usepackage{ACL2023}

\usepackage{times}
\usepackage{latexsym}

\usepackage[T1]{fontenc}

\usepackage[utf8]{inputenc}

\usepackage{microtype}

\usepackage{inconsolata}

\usepackage{graphicx}

\usepackage{booktabs} 
\usepackage{multirow} 
\usepackage{amsmath} 
\usepackage{amssymb}
\usepackage{tikz}
\usepackage{pgfplots}
\usepackage{pgf-pie}
\usepackage{subfigure}
\usepackage{amsfonts}
\usepackage{pifont}
\newcommand{\cmark}{\ding{51}}%
\newcommand{\xmark}{\ding{55}}%
\definecolor{category1}{RGB}{255,153,136}
\definecolor{category2}{RGB}{247,196,127}
\definecolor{category3}{RGB}{255,227,148}
\pgfplotsset{compat=1.18} 
\title{The Music Maestro or The Musically Challenged, A Massive Music Evaluation Benchmark for Large Language Models}

\author{Jiajia Li$^{1,2}$, Lu Yang$^{3}$, Mingni Tang$^{3}$, Cong Chen$^{4}$, Zuchao Li$^{3,}$\thanks{$\ $  Corresponding author. This work was supported by the National Natural Science Foundation of China (No. 62306216, No. 72074171 and No. 72374161), the Natural Science Foundation of Hubei Province of China (No. 2023AFB816).}, Ping Wang$^{1,2,*}$, Hai Zhao$^{5}$\\
$^{1}$School of Information Management, Wuhan University, Wuhan, China\\
$^{2}$Key Laboratory of Archival Intelligent Development and Service, NAAC \\
$^{3}$School of Computer Science, Wuhan University, Wuhan, China \\
$^{4}$School of Music, Shenyang Conservatory of Music, Shenyang, China\\ 
$^{5}$Department of Computer Science and Engineering, Shanghai Jiao Tong University\\
{\tt \{cantata, yang\_lu, minnie-tang, zcli-charlie, wangping\}@whu.edu.cn}
}
\begin{document}
\maketitle
\begin{abstract}
Benchmark plays a pivotal role in assessing the advancements of large language models (LLMs). While numerous benchmarks have been proposed to evaluate LLMs' capabilities, there is a notable absence of a dedicated benchmark for assessing their musical abilities. To address this gap, we present ZIQI-Eval, a comprehensive and large-scale music benchmark specifically designed to evaluate the music-related capabilities of LLMs.
ZIQI-Eval encompasses a wide range of questions, covering 10 major categories and 56 subcategories, resulting in over 14,000 meticulously curated data entries. By leveraging ZIQI-Eval, we conduct a comprehensive evaluation over 16 LLMs to evaluate and analyze LLMs' performance in the domain of music.
Results indicate that all LLMs perform poorly on the ZIQI-Eval benchmark, suggesting significant room for improvement in their musical capabilities.
With ZIQI-Eval, we aim to provide a standardized and robust evaluation framework that facilitates a comprehensive assessment of LLMs' music-related abilities. The dataset is available at GitHub\footnote{https://github.com/zcli-charlie/ZIQI-Eval} and HuggingFace\footnote{https://huggingface.co/datasets/MYTH-Lab/ZIQI-Eval}.
\end{abstract}

\section{Introduction}
In recent years, large language models (LLMs) have made significant advancements, revolutionizing various natural language processing tasks~\cite{pami}. 
These models have showcased their proficiency in tasks such as accessing and reasoning about world knowledge. 

Benchmark evaluation has played a crucial role in assessing and quantifying the performance of LLMs across different domains.
Specific benchmarks tailored to particular tasks such as coding~\citep{program}, reading comprehension~\citep{multispanqa}, and mathematical reasoning~\citep{cobbe2021training}, in light of the advancements made by LLMs, are increasingly regarded as inadequate for assessing their comprehensive capabilities. Consequently, there has been a surge in the emergence of more comprehensive benchmarks~\citep{liang2022holistic,srivastava2022beyond}. 

However, both the specific and comprehensive benchmarks have failed to adequately address the musical capability of large language models. Music is an essential part of human life and culture, and assessing LLMs' comprehension and generation of music presents a unique and challenging task. This oversight emphasizes the necessity for a comprehensive evaluation framework specifically designed to capture the nuances of the musical domain.

Therefore, we present ZIQI-Eval, an extensive and comprehensive music benchmark specifically crafted to assess the music-related abilities of LLMs. ZIQI-Eval comprises a diverse range of questions, systematically organized into 10 major categories and 56 subcategories. These categories cover various aspects of music, including music theory, composition, genres, instruments, and historical context. 
In addition, this music benchmark actively contributes to the recognition of female music composers. By incorporating valuable content from these composers, it rectifies the gender disparity prevalent in historical literature, fostering advancement and inclusivity within the realm of music scholarship.
With over 14,000 carefully crafted data entries, ZIQI-Eval provides a rich and extensive resource for evaluating LLMs' comprehension and generation of music-related content.

Utilizing ZIQI-Eval, we conducted a comprehensive experiment over 16 LLMs, comprising API-based models and open-source models, to evaluate the performance of LLMs in the realm of music.
Specifically, we fed music knowledge or the first half of a musical score, along with four options, to the LLMs to assess their ability to select the correct option and provide meaningful explanations. 
With an average F1 of just 58.68, even the top-performing model, GPT-4, falls short in demonstrating comprehensive music understanding and generation capabilities. 
This observation not only exposes the overlooked aspect of music in LLMs but also emphasizes the significance of ZIQI-Eval in bridging this gap and tackling the inherent challenges associated with it.

\textbf{Our Contribution} \textit{\textbf{(1)}} We find that existing evaluations of the capabilities of LLMs have overlooked their musical abilities. Therefore, we propose ZIQI-Eval benchmark, a manually curated, large-scale, and comprehensive benchmark for evaluating music-related capabilities. It consists of 10 major categories and 56 subcategories, encompassing over 14,000 data entries. \textit{\textbf{(2)}} We conduct evaluations on the music comprehension and music generation capabilities of 16 LLMs and find that almost all of them struggle to understand music effectively, let alone generate it. \textit{\textbf{(3)}} We explore the issue of bias in LLMs' music capabilities, focusing on gender bias, racial bias, and region bias. Analysis indicates that over 35\% of LLMs exhibit biases, with region bias being the most severe. 

\section{Related Work}
\paragraph{\textbf{Music Comprehension}}
Inspired by the field of natural language processing (NLP), previous studies represented music as embedding sequences for music understanding.
\citet{chuan} and \citet{pirhdy} partition music pieces into distinct, non-overlapping segments of fixed duration, and train embeddings for each segment.

Later, with the development of LLMs, recent research has utilized the modeling capabilities of these models to further enhance the understanding of music~\citep{mert}.
MidiBERT~\citep{midibert} and MusicBERT~\citep{musicbert} both utilize pre-trained BERT to tackle symbolic-domain discriminative music understanding tasks. 
MusicBERT further designs OctupleMIDI encoding and bar-level masking strategy to enhance pre-training with symbolic music data.
\citet{llark} extracts music-related information from an open-source music dataset and uses instruction-tuning to instruct their proposed model LLark to do music understanding, music captioning, and music reasoning. 
NG-Midiformer~\citep{ngramtian} first processes music pieces into sequences, followed by leveraging N-gram encoder to understand symbolic music.

CLaMP~\citep{clamp} retrieves symbolic music information through learning cross-modal representations between natural language and symbolic music.

\paragraph{\textbf{Music Generation}}
Before the proliferation of LLMs, there are some other traditional methods proposed for music generation, mainly falling into three categories: neural networks, neural audio codecs, and diffusion models.

~\citet{gansynth},~\citet{gacela},~\citet{greshler}
employ neural network architectures such as CNNs, RNNs, or GANs to achieve music generation.
A neural audio codec typically contains an encoder and a decoder.~\citet{musae} follows the typical structure.
Some models such as Jukebox~\citep{jukebox}, and MusicLM~\citep{musiclm} further insert a vector quantizer between the encoder and the decoder to learn a discrete latent representation.
A diffusion model iteratively adds Gaussian noise and then learns to reverse the diffusion process to construct desired data samples from the noise. ~\citet{diffwave} proposes DiffWave, a non-autoregressive model that converts the white noise signal into structured waveform through a Markov chain. 
~\citet{diffsound} utilizes latent diffusion approach to generate high-quality music.

Since the advent of LLMs, researchers gradually began to explore the application of LLMs in music domain~\citep{DBLP:conf/aaai/0001WZLZZZ20,DBLP:conf/aaai/LiGCS23, musecoco, chatmusician, musicllama}.
AudioGen~\citep{audiogen} and MusicGen~\citep{musicgen} both use an autoregressive transformer-based decoder~\citep{transformer} that operates on the discrete audio tokens. Macaw-LLM~\citep{macawllm} incorporates visual, audio, and textual information by using an alignment module to unite multi-modal features to textual features for LLM to generate response. 
Some models~\citep{musilingo,m2ugen} exploit the potential of LLM to bridge multi-modal music comprehension and generation.

\begin{figure*}[t]
    \centering
\includegraphics[width=\textwidth]{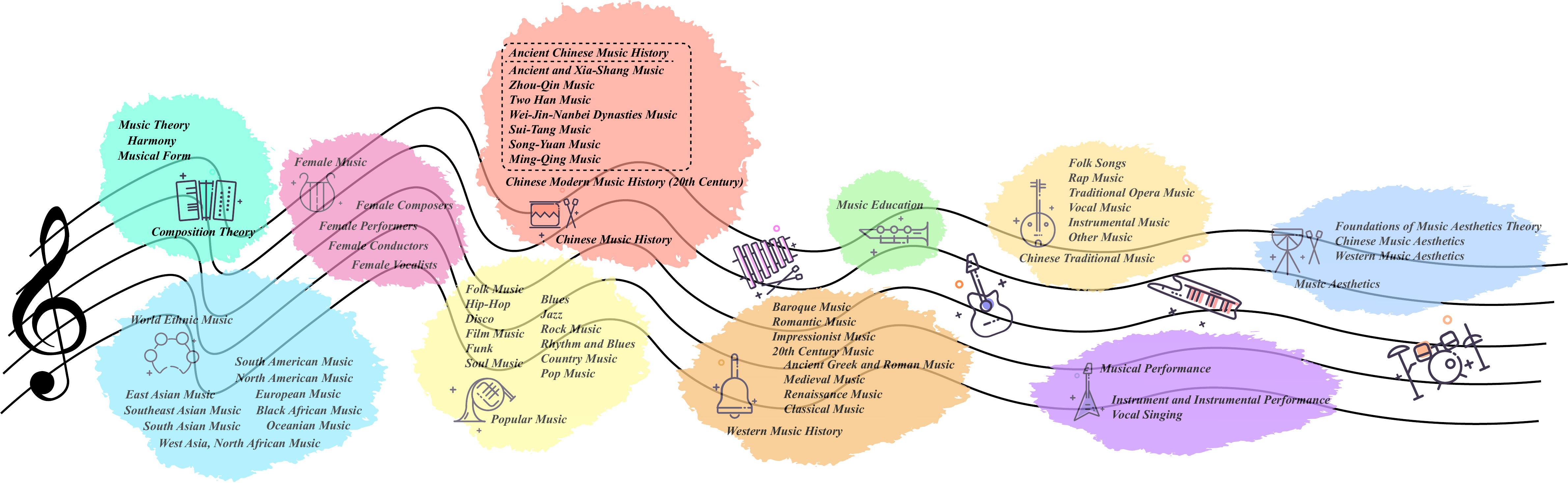}
    \caption{
        ZIQI-Eval task overview.
    }
    \label{fig:task-overview}
\end{figure*}

\paragraph{\textbf{Benchmark Evaluations}}
Benchmark evaluation plays a crucial role in assessing the development of LLMs. Previous specific benchmarking efforts~\citep{math,commonsense,arcmmlu} focused on evaluating certain capabilities of models in individual tasks or single-task types. However, with the advancement of LLMs, these benchmarks have become insufficient for comprehensive and accurate assessment of LLM capabilities. Consequently, researchers have proposed more comprehensive and challenging benchmarks~\citep{mmlu,cmmlu,ceval} to test whether LLMs possess general world knowledge and reasoning ability. Additionally, there are task-specific evaluations such as LawBench~\citep{lawbench} and ArcMMLU~\citep{arcmmlu}. However, whether in English or Chinese, there is currently only a few benchmarks in music domain~\citep{marble, ten}, let alone for evaluating the musical abilities of LLMs, despite music being an important part of human life. Therefore, we propose ZIQI-EVAL, a benchmark for evaluating the musical abilities of LLMs, to fill the gap in benchmark evaluations of LLMs' musical capabilities.

\section{ZIQI-Eval Benchmark}
\subsection{Dataset Curation}
\paragraph{\textbf{General Principle}} 
This dataset integrates the renowned music literature database Répertoire International de Littérature Musicale (RILM), providing a broad research perspective and profound academic insights. The inclusion of "The New Grove Dictionary of Music and Musicians" injects the essence of musical humanism into the dataset. Furthermore, dozens of domestic and foreign monographs, such as "Music in Western Civilization" by Paul Henry Lang, the availability of past exam materials from Baidu Wenku, and the advanced data processing capabilities of GPT-4~\citep{gpt-4}, collectively enhance the data integrity and reliability of the benchmark.

    
\paragraph{\textbf{Data Statistics}}
ZIQI-Eval dataset consists of two parts: music comprehension question bank and music generation question bank.

The music comprehension question bank which is presented in the form of multiple-choice questions consists of 10 major categories and 56 subcategories, encompassing 14244 data entries. It not only includes traditional classifications such as music performance, composition theory, and world ethnic music, but also covers popular music, Western music history, Chinese music history, Chinese traditional music, music aesthetics, and music education. The topics range from popular music, rock music, blues, to female music and more. Additionally, the dataset adopts a decentralized design philosophy, fully showcasing the diversity and inclusiveness of global music cultures. 

The music generation question bank consists of 200 questions, testing the ability of music continuation. 
Considering the difficulty in the evaluation of the generated music, the music generation questions are also presented in the form of multiple-choice questions.


\begin{figure}[t]
    \centering
\includegraphics[width=0.9\columnwidth]{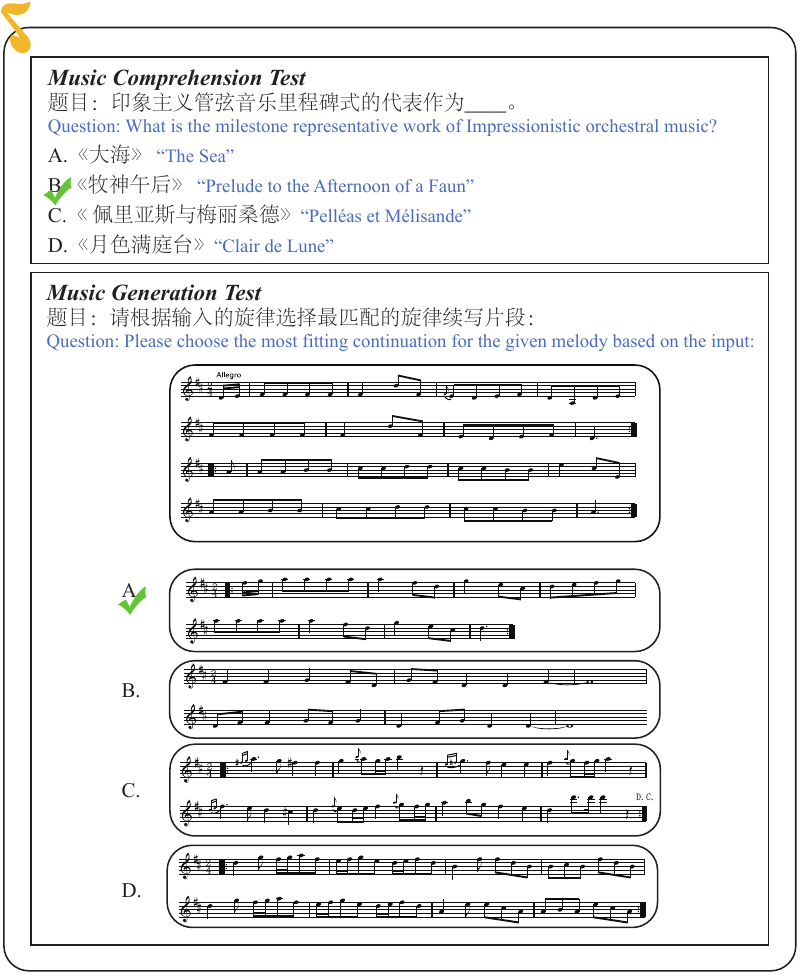}
    \caption{
        Examples of music comprehension and music generation test.
    }
    \label{fig:example}
\end{figure}

We conduct a comprehensive evaluation of LLMs' music capabilities across the entire dataset. It is worth mentioning that this music dataset has made positive contributions in highlighting female music composers. By including relevant content about female composers, it addresses the gender imbalance in historical literature and promotes progress and inclusivity in the music academic community. This initiative not only reflects the benchmark's profound recognition of gender equality issues but also demonstrates its efforts in advancing the diversification of the music field.

\subsection{Evaluation Criteria}
The evaluation is divided into two parts: music comprehension test and music generation test. The music comprehension test aims to assess the LLMs' music comprehension abilities, specifically their understanding of music harmony, melody, and rhythm. The music generation test, on the other hand, seeks to evaluate the LLMs' capacities for music generation, namely their ability to generate music across diverse styles and genres.

\paragraph{\textbf{Music Comprehension Test}}
We turn the music-related knowledge into the question stem and provide them with options to LLMs, making LLMs to choose the right answer. For example, as shown in Figure~\ref{fig:example}, take ``What is the milestone representative work of Impressionistic orchestral music?'' as the stem,  ``The Sea'', ``Prelude to the Afternoon of a Faun'', ``Pelléas et Mélisande'', and ``Clair de Lune'' as the options, we examine whether LLMs can select the right answer ``Prelude to the Afternoon of a Faun''.

\paragraph{\textbf{Music Generation Test}}


Given that most LLMs can only accept textual inputs, we utilize ABC notation to convert the musical scores of audio into a textual format, which serves as the input for LLMs.
We partition the sheet music written in ABC notation into two segments. The initial segment serves as the question, while the subsequent segment presents four alternative options, also in ABC notation, for the potential continuation of the composition. 
Then we make LLMs discern the most likely continuation fragment, assessing their music continuation ability. For instance, as shown in Figure~\ref{fig:example}, we split the original score,
and test whether LLMs have the ability to choose the most fitting option.
To facilitate understanding, we visualize the ABC notation.

\section{Experiments}
\subsection{Setup}
\paragraph{\textbf{Baselines}} 
We comprehensively assess 16 LLMs, including API-based models and open-source models. The API-based models contain GPT-4 (gpt-4-1106-preview)~\citep{gpt-4}, GPT-3.5-Turbo~\citep{chatgpt}, Claude-instant-1.2~\citep{claude}, and ERNIE-Bot~\citep{ERNIE-bot} series. The open-source models contain Aquila-7B~\citep{aquila}, Bloomz-7B~\citep{bloomz}, ChatGLM2-6B~\citep{chatglm2}, 
Mixtral-8x7B~\citep{mixtral}, XuanYuan-70B~\citep{xuanyuan}, Qwen~\citep{qwen} series, and Yi~\citep{yi} series.

\begin{table*}[htb]
 \centering
 \renewcommand\tabcolsep{8pt} 
 \scalebox{0.85}{
 \begin{tabular}{lcccccccccc}
 \toprule
  \multirow{2}{*}{Models}  & \multicolumn{3}{c}{Music Comprehension Test} & \multicolumn{3}{c}{Music Generation Test} 
    \\\cmidrule(lr){2-4}\cmidrule(lr){5-7}
             & Precision & Recall (Acc.) & F1 & Precision & Recall (Acc.)& F1  \\\midrule  
GPT-4 & 63.15 & 62.93 & 63.04 & 55.15 & 53.50 & 54.31 \\
GPT-3.5-Turbo & 55.96 & 50.18 & 52.91 & 31.77 & 30.50 & 31.12 \\
Claude-instant-1.2 & 64.20 & 45.86 & 53.50 & 25.13 & 25.00 & 25.06 \\
ERNIE-Bot & 74.91 & 49.96 & 59.94 & 30.05 & 29.00 & 29.52 \\
ERNIE-Bot-Speed & 41.81 & 31.18 & 35.72 & 42.00 & 42.00 & 42.00 \\
ERNIE-Bot-Turbo & 50.90 & 47.88 & 49.34 & 25.50 & 25.50 & 25.50 \\
ERNIE-Bot-8k & 53.62 & 53.17 & 53.39 & 30.11 & 26.50 & 28.19 \\
 \midrule
Aquila-7B & 46.57 & 29.06 & 35.79 & 22.50 & 9.00 & 12.86 \\
Bloomz-7B & 35.11 & 31.97 & 33.47 & 29.46 & 19.00 & 23.10 \\
ChatGLM2-6B & 65.80 & 39.82 & 49.61 & 24.80 & 15.50 & 19.08 \\
Mixtral-8x7B & 43.58 & 43.39 & 43.49 & 31.00 & 31.00 & 31.00 \\
XuanYuan-70B & 80.73 & 37.70 & 51.40 & 23.60 & 21.00 & 22.22 \\
        Qwen-7B & 35.95 & 14.54 & 20.70 & 24.05 & 9.74 & 13.87 \\
       Qwen-14B & 30.04 & 17.98 & 22.49& 23.66 & 15.50 & 18.73  \\
    Yi-6B & 60.00& 11.06 &18.68 & 0.00 & 0.00 & 0.00\\
        Yi-34B &  32.24& 16.76 & 22.06& 12.12 & 2.00 & 3.43 \\
    \bottomrule
 \end{tabular}
 }
 \caption{Main results(\%) of the Music Comprehension Test and Music Generation Test in ZIQI-Eval. Part 1: API-based models; Part 2: Open-source models. 
 }
 \label{tab:mainres}
 \vspace{1mm}
\end{table*}

\paragraph{\textbf{Metrics}}
We use a regular expression $R$, namely $r'[ABCD]'$, to match the answer and consider the first uppercase letter $\in  \left \{ \text{`A'}, \text{`B'}, \text{`C'}, \text{`D'} \right \} $ matched as the response.
We define Accuracy (Acc.) as the proportion of correctly answered questions among all questions. Precision is the proportion of correctly answered questions among the questions predicted as A/B/C/D. Recall is the proportion of correctly answered questions among the total number of questions that should be answered as A/B/C/D. In this case, the total number of questions that should be answered as A/B/C/D is actually the total number of questions, so the recall metric is equivalent to the accuracy metric. F1 score is the weighted harmonic mean of precision and recall. The specific formulas for these metrics are as follows:
\begin{gather*}
\tilde{X} = G\left ( X \right ) \\
\hat{y} = R\left ( \tilde{X} \right )  \\
    Precision = \frac{\sum_{i=1}^{N} \mathbb{I}\left ( \hat{y_i} = y_i \right ) }{V} \\
    Recall (Acc.) = \frac{\sum_{i=1}^{N} \mathbb{I}\left ( \hat{y_i} = y_i \right ) }{N} \\
    F1 = \frac{2\ast Precision\ast Recall}{Precision + Recall} 
\end{gather*}
where $G\left ( \cdot  \right )$ is the LLM generation process, $\tilde{X}$ is the generated string, $R\left ( \cdot  \right ) $ is applying the regular expression for answer retrieval, $\hat{y}$ is the predicted answer, $V$ is the number of questions predicted as A/B/C/D, $N$ is the total number of the questions, and $\mathbb{I}\left ( \cdot  \right )$ is the indicator function.

\subsection{Results}
Table~\ref{tab:mainres} presents the main results of ZIQI-Eval. Based on the results, we can find that:

\paragraph{\textbf{\uppercase\expandafter{\romannumeral1}.}} \textbf{Overall, the performance of all LLMs on the ZIQI-Eval benchmark is poor.} 
In both music comprehension test and music generation test, the majority of LLMs have not surpassed the passing threshold of 60. Their F1 scores generally hover between 30 and 50, performing only marginally better than random selection. Even the top-performing model, GPT-4, achieved F1 of only 63.04 and 54.31 in the respective tests. This glaring discrepancy highlights the inadequate consideration given to music abilities within current LLM models and underscores the formidable challenges posed by the ZIQI-Eval benchmark.

\paragraph{\textbf{\uppercase\expandafter{\romannumeral2}.}} \textbf{API-based models perform better than open-source models.} In the evaluation of music comprehension test, API-based models generally exhibit higher F1 compared to open-source models. The F1 of API-based models is basically distributed between 50 and 70, while open-source models mostly range between 20 and 50. Only specific open-source models like XuanYuan-70B can achieve an F1 higher than 50.

\begin{figure*}[htb]
  \centering
  \setlength{\abovecaptionskip}{0cm}
  \setlength{\belowcaptionskip}{-0.3cm} 
  \scalebox{1.0}{
  \subfigure[F1]{
    \label{F1} 
    \includegraphics[width=0.64\textwidth]{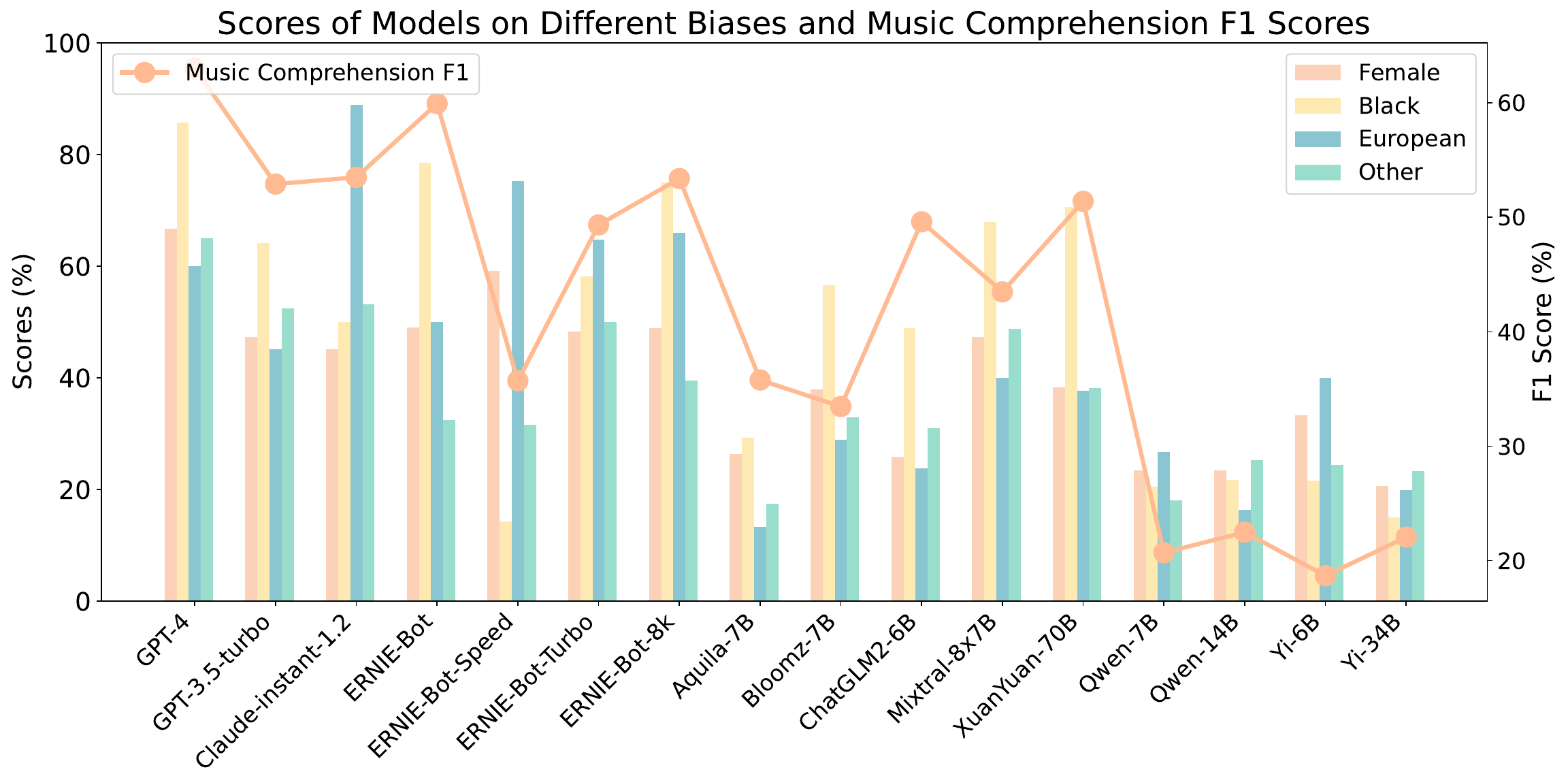}}
  \subfigure[Proportion]{
    \label{proportion} 
    \includegraphics[width=0.36\textwidth]{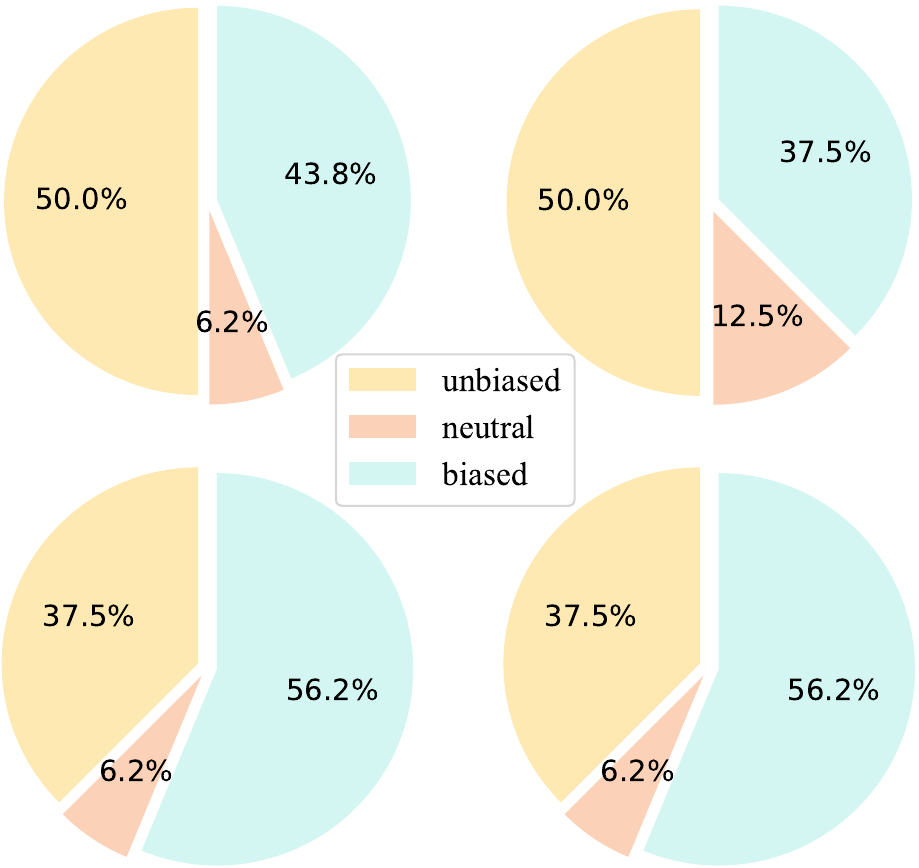}}}
  \caption{Performance of LLMs on gender bias, racial and region bias. Subfigure (a) shows F1 scores of every LLM regarding biases. A line graph is plotted using the average F1 of each LLM to show LLM's overall bias condition. Subfigure (b) depicts the distribution of the biases. Top left: gender bias, top right: racial bias, bottom left: European region, bottom right: other regions.}
  \label{bias} 
\end{figure*}

In the evaluation of music generation questions, API-based models apparently outperform open-source models with significantly higher F1. The highest F1 achieved by an API-based model is 54.31, surpassing the highest F1 of 31.00 in open-source models.

\paragraph{\textbf{\uppercase\expandafter{\romannumeral3}.}} \textbf{The music capabilities of LLMs are dependent but not solely on parameter scale.} There is a certain degree of relationship between the musical ability and parameter scale of LLM models within the same series, while the musical ability of LLM models from different series is not strongly correlated with parameter scale.

The Qwen series and Yi series LLMs consistently show improvements in both music comprehension and generation F1.
Contrary to expectations, the model with significantly different parameter scales, ChatGLM2-6B and Yi-34B, exhibits higher F1 in music comprehension for the ChatGLM2-6B model, surpassing Yi-34B by 27.55. Even among models with similar parameter scales, there can be considerable differences in performance. For example, the Yi-6B model achieves a music comprehension F1 of only 18.68, while ChatGLM2-6B achieves an F1 of 49.61, resulting in a significant difference of 30.93 between the two F1 scores.

\paragraph{\textbf{\uppercase\expandafter{\romannumeral4}.}} \textbf{The instruction-following abilities of LLMs are not directly linked to their music capabilities.} The precision scores of LLMs are strongly correlated with their instruction-following abilities. However, a strong instruction-following capability does not necessarily indicate strong musical capabilities in LLMs. Some LLMs may score highly in terms of precision, but they struggle to effectively comprehend and generate music. Yi-6B serves as a clear example where the comprehension precision score reaches 60.00, but the F1 is only equivalent to 18.68.

\paragraph{\textbf{\uppercase\expandafter{\romannumeral5}.}} \textbf{The music generation capabilities of LLMs are in need of improvement.} Even though some LLMs demonstrate a decent understanding of music, their music generation capabilities still have room for improvement. In general, the F1 for music generation test in LLMs are lower compared to music comprehension test. The difference can be quite significant, such as ERNIE-Bot-8k, where the score for music comprehension test is higher by 25.20 compared to music generation test.

\begin{table}[htb]
 \centering
 \renewcommand\tabcolsep{2pt} 
 \scalebox{0.75}{
     \begin{tabular}{lcccc}
     \toprule
      \multirow{2}{*}{Models}  & \multirow{2}{*}{Female} & \multirow{2}{*}{Black} & \multicolumn{2}{c}{Region} \\
        \cmidrule(lr){4-5}
                 & & & European & Other \\
     \midrule  
     GPT-4                   & 66.67   & 85.71   & 60.00   & 64.97   \\
GPT-3.5-turbo           & 47.37   & 64.15   & 45.09   & 52.49   \\
Claude-instant-1.2      & 45.16   & 50.00   & 88.89   & 53.13   \\
ERNIE-Bot & 49.04 & 78.57 & 49.96& 32.48\\
ERNIE-Bot-Speed         & 59.09   & 14.29   & 75.28   & 31.61   \\
ERNIE-Bot-Turbo         & 48.26   & 58.18   & 64.80   & 50.03   \\
ERNIE-Bot-8k            & 48.95   & 75.00   & 65.92   & 39.49   \\
     \midrule
Aquila-7B               & 26.31   & 29.27   & 13.25   & 17.45   \\
Bloomz-7B             & 37.87   & 56.60   & 28.90   & 32.94   \\
ChatGLM2-6B             & 25.88   & 48.89   & 23.84   & 30.95   \\
Mixtral-8x7B            & 47.32   & 67.86   & 40.00   & 48.76   \\
XuanYuan-70B &38.35 & 70.54& 37.70  &38.17\\
Qwen-7B & 23.40 & 20.51 & 26.67 & 18.11 \\
            Qwen-14B & 23.45 & 21.74 &16.30 & 25.22\\
           Yi-6B &33.26 & 21.62 & 40.00 & 24.39\\
           Yi-34B & 20.63 & 15.00 & 19.85 & 23.29 \\
     \midrule
            Average &40.06   & 48.62   & 43.53   & 36.47\\
     \bottomrule
     \end{tabular}
    }
 \caption{Results(\%) of Female Music F1, Black African Music F1, F1 of European Music and other regions' music in World Ethnic Music. 
 }
 \label{tab:ablation}
 \vspace{1mm}
\end{table}
\section{Analysis}
In addition to the overall evaluation of LLMs on the dataset, we are also interested in the models' ability for specific categories.

\subsection{Does LLM show any bias towards questions related to women?}
\label{gender}
We compare the F1 of LLMs in the female music theme with the average F1 obtained by LLMs in the female music theme to analyze whether there is bias in LLMs towards female music. We categorize LLMs into three groups: LLMs without gender bias (above the average F1), LLMs with no significant bias (deviating within a range of $\pm$5.0\% from the average F1), and LLMs with gender bias (below the average F1).

Because we do not fine-tune LLMs, the results reflect the inherent biases of the LLMs themselves. According to the results of Table~\ref{tab:ablation}, 50.00\% of the models have no gender bias, 6.25\% of the models are neutral or have no significant bias, and 43.75\% of the models have gender bias.
LLMs with F1 lower than the average F1 tend to overlook relevant content related to female music themes. Yi-34B and Aquila-7B, in particular, have significantly lower scores than the average, indicating a notable gender bias issue in these two models.

\subsection{Does LLM exhibit bias toward different races?}
We calculate the F1 of LLMs for the subtopic of Black African music, using the same partitioning method as for determining gender bias, to assess whether there is racial bias in LLMs.
According to the results of Table~\ref{tab:ablation}, 50.00\% of the models have no racial bias, 12.50\% of the models are neutral or have no significant bias, and 37.50\% of the models have racial bias.
The F1 of ERNIE-Bot-Speed and Yi-34B are below the mean by 34.33 and 33.62 respectively, indicating a significant racial bias in these two models.

\subsection{Does LLM display bias in terms of region?}
We seek to investigate whether LLMs are influenced by Eurocentrism, which positions Europe as the cultural and knowledge center, potentially leading to lower evaluations or neglect of contributions from non-central regions and resulting in biases against these regions. To assess the presence of region bias, we computed the F1 for the European Music subtheme within World Ethic Music, and the average F1 for other subthemes within World Ethic Music.
Among the LLMs, 43.75\% exhibited higher F1 rates in European Music compared to other regional music, while 37.5\% of LLMs demonstrated higher F1 rates in European Music than the average F1 rate within European Music. These findings suggest that LLMs are influenced by Eurocentrism and exhibit bias towards non-central regions.
Surprisingly, ERNIE-Bot-Speed and Claude-instant-1.2 exhibit significantly higher F1 scores in European music compared to other regions, 43.67 and 35.76 respectively, demonstrating a clear regional bias inclination.

From Figure~\ref{proportion}, it is evident that LLMs demonstrate similar tendencies towards gender bias, racial bias, and region bias, displaying a trend where both ends (with bias and without bias) are relatively higher, while the middle (neutral) is lower. Some LLMs have F1 scores significantly lower than the mean, such as Yi-34B with an F1 lower than the mean by over 50\%, suggesting that LLMs still have a long way to go in eliminating biases, as shown in Figure~\ref{F1}. It is worth noting that LLMs with a propensity for gender bias are likely to exhibit racial bias and region bias as well, as evidenced by models such as Aquila-7B, Qwen series, and Yi series. Consequently, it is imperative for future developments in LLMs to address biases comprehensively, not limited to gender, racial and region biases.

\section{Futher Analysis}
\subsection{Phenomenon Analysis of LLMs }
To further explore the generation capabilities of LLMs in the realm of music, we conducted an in-depth analysis of the responses provided by each model. Our findings categorize the existing LLMs into three distinct types:

\paragraph{\textbf{\uppercase\expandafter{\romannumeral1}.}} \textbf{Lack of melodic understanding: }This type includes LLMs that demonstrate a complete lack of comprehension regarding musical notation. When faced with questions that require the continuation of a melody after a format transformation, these models predominantly resort to evasion, often responding with statements like "Unable to determine, need more information." They fail even to understand the format of the input melody. ChatGLM2-6B and Aquila-7B are prototypical examples of this type, characterized by a high frequency of evasive responses, resulting in a significantly low efficacy in their replies. A notable phenomenon is their tendency to "guess" by consistently selecting option A, leading to most responses without any analytical explanation. For instance, in the responses from ChatGLM2-6B, option A was chosen up to 60\%. Besides a preference for option A, Aquila-7B also shows a partiality towards option D.

\paragraph{\textbf{\uppercase\expandafter{\romannumeral2}.}} \textbf{Limited appreciation, misaligned with human preferences: }A representative model in this type is ERNIE-Bot-8K. This model provides highly interpretable analyses for each option of every question, offering seemingly logical explanations concerning melody, rhythm, and pitch. However, the model's performance, with F1 barely exceeding that of random selection, underscores the challenge of encapsulating the subjective essence of music appreciation through algorithmic processes. This discrepancy not only highlights the limitations of current AI models in understanding complex, subjective domains but also underscores the need for more sophisticated approaches that can better capture the intricacies of human preferences. 

\begin{table*}[]
    \centering
\resizebox{\textwidth}{!}{\begin{tabular}{lcccc}
\toprule
  Educational Qualifications and Major & Music Major? & With Music Background? & Score Ranges & Average Score (Precision) \\
  \midrule
  High-school & \textcolor{red}{\xmark} & \textcolor{red}{\xmark} & 21 - 51 & 34.50 \\
  Undergraduate & \textcolor{red}{\xmark} & \textcolor{red}{\xmark} & 26 - 62 & 39.91 \\
  Master & \textcolor{red}{\xmark} & \textcolor{red}{\xmark} & 29 - 60 & 45.80 \\
  Ph.D. & \textcolor{red}{\xmark} & \textcolor{red}{\xmark} & 26 - 51 & 44.00 \\
  Undergraduate & \textcolor{red}{\xmark} & \color[HTML]{008114}\cmark & 27 - 70 & 45.83 \\
  Master & \textcolor{red}{\xmark} & \color[HTML]{008114}\cmark & 51 - 65 & 57.75 \\
  Ph.D. & \textcolor{red}{\xmark} & \color[HTML]{008114}\cmark & 64 & 64.00 \\
  High-school & \color[HTML]{008114}\cmark & \color[HTML]{008114}\cmark & 26 - 71 & 37.81 \\
  Undergraduate & \color[HTML]{008114}\cmark & \color[HTML]{008114}\cmark & 30 - 88 & 51.52 \\
  Master & \color[HTML]{008114}\cmark & \color[HTML]{008114}\cmark & 35 - 89 & 64.75 \\
  Ph.D. & \color[HTML]{008114}\cmark & \color[HTML]{008114}\cmark & 28 - 88 & 64.91 \\
  GPT-4 & - & - & - & 67.54 \\\bottomrule
\end{tabular}}
    \caption{Comparison between GPT-4 and humans with different musical backgrounds and educational qualifications.}
    \label{tab:humanvsgpt}
\end{table*}
\begin{table}
    \centering
   \resizebox{0.45\textwidth}{!}{ \begin{tabular}{lcccccccccccccc}
    \toprule
  & master & Ph.D. & GPT-4 \\
  \midrule
  Western Music History & 66.28 & 64.20 & 75.00 \\
  Popular Music & 54.78 & 50.00 & 81.82 \\
  World Ethnic Music & 48.54 & 51.52 & 61.11 \\
  Chinese Traditional Music & 61.62 & 68.94 & 50.00 \\
  Chinese Music History & 81.38 & 79.72 & 71.43 \\
  Female Music & 53.29 & 52.27 & 75.00 \\
  Black African Music & 82.89 & 72.73 & 100.00 \\
  Musical Performance & 78.29 & 68.18 & 100.00 \\
  Music Education & 52.63 & 63.64 & 50.00 \\
  Music Aesthetics & 42.54 & 51.52 & 50.00 \\
  Composition Theory & 63.16 & 67.53 & 50.00 \\
  Film Music & 46.05 & 27.27 & 100.00 \\
  Music Generation & 7.89 & 18.18 & 25.00 \\
\bottomrule
\end{tabular}
}
\caption{Comparison of performance between highly educated individuals and GPT-4 in each category.}
\label{highly-educated}
\end{table}

\paragraph{\textbf{\uppercase\expandafter{\romannumeral3}.}} \textbf{Relatively good appreciation skills: }GPT-4 stands out as a typical example of this type. Its responses consider aspects such as melodic coherence, stylistic similarity, and the seamless integration of musical structures, aligning to a certain extent with human preferences. Further analysis of the questions GPT-4 answered reveals a incorrectly strong inclination towards musical continuity. In many instances, it is observed that GPT-4 prioritized coherence, which leads to the selection of incorrect options.

\subsection{Analysis of GPT-4}

Taking GPT-4 as a case study, we have gained further insights into the performance of LLMs in the realm of music. The performance of GPT-4 in the domains of female music and world ethnic music indicates a commendable understanding of specific musical areas, reflecting GPT-4's focus on diversity and inclusivity. 

GPT-4 has demonstrated exceptional performance in the realm of popular music, achieving scores close to 90. This may be due to the abundant and accessible resources in popular music, including lyrics, genres, and artist information. The popularity and media coverage of pop music may also have facilitated the model's learning efficiency in this field.

It has also scored high in western music history and musical performance, showcasing its capability in processing music history and practical music-making. The higher scores in western music history over all other regions suggest a certain degree of region bias.

In Chinese music history and Chinese traditional music, GPT-4 demonstrates relatively low performance, revealing its deficiencies in handling Chinese musical content. 

In the area of music aesthetics, GPT-4 scored low, revealing a significant weakness. This may be attributed to the complexity and subjectivity of music aesthetics, which might surpass the model's ability to learn from existing textual materials, indicating that there is room for improvement in the model's perception, evaluation, and theoretical analysis of music.

Through analysis, we identify that GPT-4 tends to make errors in several distinct categories, primarily falling into three types:

\textbf{Matching Errors:} This category encompasses questions related to musical knowledge, specifically matching-type queries, such as identifying the first Hungarian national opera or the composer of "The Song of the Red Flag". GPT-4's responses often affirmatively state incorrect options, indicating inaccuracies within its knowledge base for specific factual information.

\textbf{Comprehension Errors:} These errors involve understanding specific musical terminologies and the relationships between certain concepts. Questions like "What function of art does edutainment refer to?"  or "What role do work songs play in labor as a genre of folk music?" exemplify where GPT-4 misinterprets multiple word meanings, leading to a misunderstanding of the intended concept. This suggests a need for improvement in GPT-4's understanding within the musical domain.

\textbf{Reasoning Errors: }In instances where GPT-4 correctly understands the question and possesses the relevant knowledge background, errors occur during the reasoning or calculation process, resulting in incorrect conclusions. An example can be seen in questions involving the calculation of musical intervals, where GPT-4 confuses semitones and whole tones. This indicates a gap in GPT-4's ability to perform downstream tasks that require precise musical logical deductions.

\section{Human Experiments}
We also conduct relevant user studies, and we choose the Precision metric to compare the performance of SOTA LLM (GPT-4) and humans. Due to time and cost limitations, we randomly selected 2 questions for each subcategory (including music comprehension questions and music generation questions), totaling 114 questions. We distributed surveys and recruited 165 individuals with and without a musical background, of different educational qualifications, to answer the questions. Among them, 32.1\% are non-music majors without any musical background, 8.5\% are non-music majors with some musical background, and 59.4\% are music majors with a musical background. As for educational qualifications, 21.2\% have a high school diploma, 31.5\% have a bachelor's degree, 35.2\% have a master's degree or are currently pursuing one, and 12.1\% have a doctorate or are currently pursuing one. The results can be seen in Table~\ref{tab:humanvsgpt} and Table~\ref{highly-educated}.
It can be observed that:
\paragraph{\textbf{\uppercase\expandafter{\romannumeral1}.}} 
\textbf{Individuals with a musical background tend to score higher. }
Specifically, the order of scores is as follows: Music majors with a musical background > Non-music majors with some musical background > Non-music majors without a musical background.
\paragraph{\textbf{\uppercase\expandafter{\romannumeral2}.}} 
\textbf{Individuals with higher educational qualifications tend to score higher.} Specifically, the order of scores is as follows:
Individuals with a doctorate degree or currently pursuing a doctorate > Individuals with a master's degree or currently pursuing a master's > Undergraduate students > High school students.
\paragraph{\textbf{\uppercase\expandafter{\romannumeral3}.}} 
\textbf{LLMs have an advantage in terms of knowledge breadth, while music master's and doctoral students have an advantage in their specialized domain knowledge.}
GPT-4 demonstrates music abilities that are equivalent to, or even surpass, the average level of a music doctoral student (including those currently pursuing a doctorate). This clearly proves the exceptional music capabilities of LLMs. 
In this regard, the remarkable abilities of LLMs are highly commendable, and it is believed that they will continue to bring us more surprises in the future. 
However, on the other hand, in certain knowledge domains (such as Chinese Traditional Music, Chinese Music History, Music Education, Composition Theory etc.), it is evident that the scores of music master's or doctoral students are higher. This may be due to their focused and in-depth research in a specific field during the master's and doctoral stages. 

\section{Conclusion and Future Work}
Our research sheds light on the oversight of existing evaluations in recognizing the musical abilities of large models. To address this gap, we introduce ZIQI-Eval, a comprehensive benchmark that encompasses 10 major categories and 56 subcategories, comprising over 14,000 data entries. Notably, this benchmark also actively contributes to the acknowledgment of female music composers, rectifying the gender disparity and promoting inclusivity.
We conducted a comprehensive experiment involving 16 LLMs, including both API-based and open-source models, to assess their performance in the domain of music. The results indicate that there is significant scope for enhancing the musical capabilities of existing LLMs.
We intend to create a multimodal benchmark to evaluate the musical expertise of LLMs in the future.
\section*{Acknowledgement}
Jiajia Li, Lu Yang and Mingni Tang made equal contributions to this paper. We are appreciative of the anonymous reviewers for their helpful suggestions that have improved the quality of our paper.
\section*{Limitations}
Our research to date has been exclusively focused on objective questions, without delving into the study of subjective questions. One limitation of our current music benchmark is the absence of multi-modal data. While the benchmark may excel in evaluating and comparing the quality and creativity of musical compositions based on audio data alone, it fails to incorporate other essential aspects of the music experience, such as visual elements or textual information.




\bibliography{acl2023}
\bibliographystyle{acl_natbib}

\appendix

\end{document}